# The effect of interfacial slip on the motion and deformation of a droplet in an unbounded arbitrary Stokes flow

Shubhadeep Mandal[1], Aditya Bandopadhyay[2] and Suman Chakraborty[1,2]†

[1]*Department of Mechanical Engineering, Indian Institute of Technology Kharagpur, West Bengal- 721302, India*

[2]*Advanced Technology Development Center, Indian Institute of Technology Kharagpur, West Bengal- 721302, India*

The motion and deformation of a droplet suspended in an unbounded fluid with an arbitrary, but Stokesian, imposed flow is investigated when there is a slip at the interface between the two liquids. The boundary condition at the interface is accounted by means of a simple Navier slip condition. Expressions are derived considering the effect of slip on the velocity and the shape deformation of the droplet for any arbitrary imposed flow field, and results are presented for the specific cases of shear flow and Poiseuille flow with the results of Hetstroni and Haber (*J. Fluid Mech.*, 1970, vol. 41(04), pp. 689–705); and Ramachandran and Leal (*J. Rheol.*, 2012, vol. 56(6), pp. 1555-1587) as the limiting cases of our general expressions. The modification to Faxén's law is also presented in the above perspective.

## I. INTRODUCTION

The problem of motion and deformation of a droplet or bubble suspended in a viscous liquid is of considerable interest from various perspectives of fluid mechanics such as biofluid transport, blend rheology, combustion and so on.[1,2] Active control over the generation, shape deformation and transport though devices of these soft or deformable entities is very much essential for different microfluidic processes: mixing of reagents, cell encapsulation, drug delivery, etc.[3–5] The study of the motion of droplet is a long-standing problem due to the inherent non-linearity associated with the interface deformation. Several studies over decades have shown the influence of deformability,[6–9] fluid inertia,[10–12] non-Newtonian rheology,[13–16] interfacial effects (e.g. thermocapillary effect or the presence of surfactant) [17–24] and electrical stresses [25,26] on the motion and deformation of droplets.

The paradigm of no-slip velocity at the surface of particles has been revised extensively in recent times with many proposed mechanisms accounting for a slip at the surface based on experimental and molecular dynamics evidences.[27,28] Besides, particle surfaces may be chemically treated so as to render the surface slipping relative to the suspending fluid.[29,30] Basset considered the drag acting on the particle moving due to the action of an applied force and found that the velocity of a rigid particle increases with interfacial slip until the particle starts behaving as a bubble.[31] The problem of a particle motion with interface slip demonstrates markedly altered particle velocities, and thus other bulk rheological properties.[32] The motion and dynamics of a Janus particle has also been addressed in recent times.[33] For the case of droplets and bubbles, the interface between two fluids is not always at a perfect equilibrium, leading to a relative slippage upon motion at the surface.[32,34–36] Such a slippage has been shown to significantly affect the droplet morphology which also affects the so called emulsion rheology. However, we note that so far in the existing literature, the effect of slip has been accounted only in the cases of linear flows.[24,37–39]

†Email address for correspondence: suman@mech.iitkgp.ernet.in

Stokes flow in and around droplets has a focus of current research. Despite the emergence of boundary element methods,[40] analytical solutions go a long way towards quick and accurate predictions and estimations of the motion of particles and droplets.[8,9,41,42] The general methodology proposed by Lamb,[43] and later made systematic by Happel and Brenner,[44] relies on the representation of the velocity and pressure fields by means of solid spherical harmonics. In this work, we aim at developing a general framework for a droplet suspended in another liquid with a slip at the interface for any arbitrary (yet Stokesian) flow field. We derive general expressions for the droplet velocity from a force free consideration, and then we obtain the deformed interface profile from the balance of the interfacial normal stress with the (high) surface tension. This is akin to the calculations done via the method of domain perturbation in terms of Capillary number ($Ca$), where the leading order velocity is obtained for a spherical shape and the knowledge of this field is utilized to obtain the shape deformation.

In section 2, we define the problem and discuss the pertinent boundary conditions. In section 3, we arrive at the solutions for the inner and outer velocity field and pressure fields by means of spherical harmonics - Lamb's solution. In section 4, we obtain the velocity of the droplet in this general flow condition by imposing a force-free condition. We also obtain the equation of the deformed interface for low capillary numbers. The general results obtained in the previous sections are demonstrated for simple shear flows and quadratic flows (Poiseuille flow). We present limiting cases of the general formulae presented above by validating them against various existing results.

## II. PROBLEM DEFINITION

We consider a non-neutrally buoyant droplet of viscosity $\mu_i$ and density $\rho_i$ suspended in a medium of viscosity $\mu_o$ and density $\rho_o$. The droplet is considered to be moving with yet unknown velocity $\mathbf{U}_d$. The undisturbed imposed flow-field $\mathbf{V}_\infty$ is Stokesian, but otherwise considered to be arbitrary. The fluids are assumed to be described by the Newtonian constitutive behaviour and constant density. The interface between the two fluids is assumed to be clean, i.e. free of surfactants. Moreover, the flows, both inside and outside the droplet, are assumed to be taking place at negligible Reynolds numbers, $Re \ll 1$, as a result of which the inertia plays no role in the governing equations of motion. We essentially assume that the inertia is completely negligible as a result of which the low Reynolds number corrections at $r \gg 1$(Oseen's approximation [12,45]) are neglected.

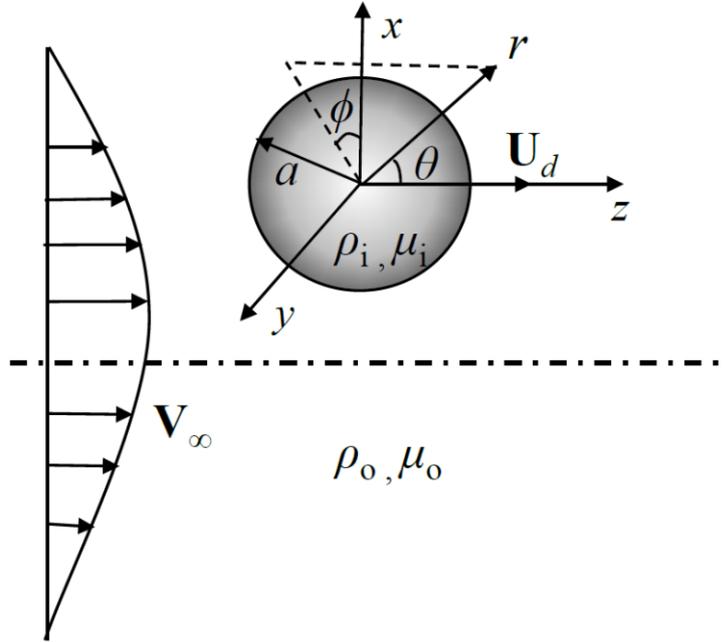

FIG. 1. The droplet moving with a velocity $\mathbf{U}_d$ with density $\rho_i$ and viscosity $\mu_i$ is spherical in its undeformed shape with a radius of $a$. The imposed flow $\mathbf{V}_\infty$ of the medium, having a density $\rho_o$ and viscosity $\mu_o$, is arbitrary. The spherical coordinate system $(r,\theta,\phi)$ is attached to the centre of the droplet.

## A. Governing equations

The governing equations for the velocity and pressure field outside the drop are the Stokes equations [46] of the form

$$\frac{1}{\mu_o}\nabla p_o = \nabla^2 \mathbf{u}_o, \ \nabla\cdot\mathbf{u}_o = 0.$$  (1)

For the velocity and pressure fields inside the droplet, we have

$$\frac{1}{\mu_i}\nabla p_i = \nabla^2 \mathbf{u}_i, \ \nabla\cdot\mathbf{u}_i = 0,$$  (2)

where the subscript $i$ and $o$ represent the field variables inside and outside the drop respectively. We note that the modified pressures $p_{i,o}$ mentioned in the above equations contain the effect of the gravity.

## B. Boundary conditions

We consider a reference frame which is moving with the droplet velocity (please see Figure 1). In the frame of the droplet, the boundary condition at infinity is represented as follows:

$$\text{at } r \to \infty, \ \mathbf{u}_o\big|_\infty = \mathbf{u}_\infty = \mathbf{V}_\infty - \mathbf{U}_d \,. \tag{3}$$

At the interface, we have the following boundary conditions:

$$\mathbf{u}_o = \mathbf{u}_i + \frac{\delta}{\mu_0}\big[\mathbf{I} - \mathbf{nn}\big] \cdot \boldsymbol{\sigma}_i \cdot \mathbf{n} \,, \tag{4}$$

$$\mathbf{u}_i \cdot \mathbf{n} = 0 \,, \tag{5}$$

$$\boldsymbol{\sigma}_o \cdot \mathbf{n} - \boldsymbol{\sigma}_i \cdot \mathbf{n} = \Gamma\left(\frac{1}{R_1} + \frac{1}{R_2}\right)\mathbf{n} \,, \tag{6}$$

where $\delta$ represents the slip parameter, $\mathbf{n}$ represents the normal vector at the interface, $\mathbf{I} - \mathbf{nn}$ represents the surface projection operator, $\boldsymbol{\sigma}_i$ and $\boldsymbol{\sigma}_o$ denote stresses inside and outside the droplet, $\Gamma$ is the coefficient of surface tension; and $R_1$ and $R_2$ denote the principle radii of the deformed droplet interface. Here, equation (3) represents the flow far away from the droplet. Equation (4) describes the matching of the tangential component of the velocity with due consideration of interfacial slip. Equation (5) states that the normal component of the velocity at the interface is zero which means the interface appears stationary with respect to the chosen reference frame. Equation (6) describes the stress balance at the interface. Equation (4) is obtained by making use of the definition of the Navier slip boundary condition [27,37,46]. The difference in the velocity in the region outside and inside the droplet at the interface is proportional to the stress at the interface; the proportionality constant, in this case, being represented by $\delta/\mu$.

It is to be noted here that the analytical determination of the velocity and pressure fields is not possible for any arbitrary deformation of interface because of the nonlinearity in the boundary conditions (equations (4)-(6)). This non-linearity is due to the fact that for problems concerning deformation of interface, the shape of the interface is not known *a priori* and has to be obtained as a part of the solution [46]. However, in the regimes of high surface tension forces relative to viscous forces, i.e, in low capillary number $Ca = \mu_0 \mathrm{V}_{\infty,\text{ref}}/\Gamma$ limits ($\mathrm{V}_{\infty,\text{ref}}$ denoting a reference velocity of the imposed flow), the droplet tends to remain spherical, thus allowing us to apply the method of domain perturbation (the leading order domain being the spherical droplet as $Ca \to 0$) to obtain the leading order velocity and pressure fields. These fields can then be utilized to obtain the higher order deformation of the droplet - a classical hallmark of the domain perturbation methods. [46] In this work, we present the first iteration of obtaining the leading order velocity and pressure field and the $O(Ca)$ interface deformation.

## III. SOLUTION METHODOLOGY

### A. Solution of the inner and outer flow fields

In this work, we utilize the general solution of Lamb in spherical harmonics to obtain the solution.[43] The outer velocity field can be written as a far field $\left( \mathbf{u}_\infty \right)$ and the harmonic field:

$$\mathbf{u}_o = \mathbf{u}_\infty + \mathbf{v}_o. \tag{7}$$

Here, $\mathbf{v}_o$ is the disturbance velocity field which decays to zero as $r \to \infty$, and is given as

$$\mathbf{v}_o = \sum_{n=1}^{\infty} \left[ \nabla \times \left( \mathbf{r} \chi_{-n-1} \right) + \nabla \Phi_{-n-1} - \frac{n-2}{2n\left(2n-1\right)\mu_o} r^2 \nabla p_{-n-1} + \frac{n+1}{n\left(2n-1\right)\mu_o} \mathbf{r} p_{-n-1} \right], \tag{8}$$

and the pressure field outside the droplet is given by

$$p_o = \sum_{n=1}^{\infty} p_{-n-1}, \tag{9}$$

where $p_{-n-1}$, $\Phi_{-n-1}$ and $\chi_{-n-1}$ represent decaying solid spherical harmonics of degree $-n-1$ of the following form

$$p_{-n-1} = \mu_o a^n r^{-n-1} \sum_{m=0}^{n} P_n^m \left( A_{-n-1}^m \cos m\phi + \hat{A}_{-n-1}^m \sin m\phi \right), \tag{10}$$

$$\Phi_{-n-1} = a^{n+2} r^{-n-1} \sum_{m=0}^{n} P_n^m \left( B_{-n-1}^m \cos m\phi + \hat{B}_{-n-1}^m \sin m\phi \right), \tag{11}$$

$$\chi_{-n-1} = a^{n+1} r^{-n-1} \sum_{m=0}^{n} P_n^m \left( C_{-n-1}^m \cos m\phi + \hat{C}_{-n-1}^m \sin m\phi \right). \tag{12}$$

Similarly the velocity and pressure fields for the inner region can be given by

$$\mathbf{u}_i = \sum_{n=1}^{\infty} \left[ \nabla \times \left( \mathbf{r} \chi_n \right) + \nabla \Phi_n + \frac{n+3}{2\left(n+1\right)\left(2n+3\right)\mu_i} r^2 \nabla p_n - \frac{n}{\left(n+1\right)\left(2n+3\right)\mu_i} \mathbf{r} p_n \right], \tag{13}$$

and

$$p_i = \sum_{n=1}^{\infty} p_n, \tag{14}$$

where $\chi_n$, $\Phi_n$ and $p_n$ represent growing solid spherical harmonics of degree $n$ of the following form

$$p_n = \mu_i a^{-n-1} r^n \sum_{m=0}^{n} P_n^m \left( A_n^m \cos m\phi + \hat{A}_n^m \sin m\phi \right), \tag{15}$$

$$\Phi_n = a^{-n+1} r^n \sum_{m=0}^{n} P_n^m \left( B_n^m \cos m\phi + \hat{B}_n^m \sin m\phi \right), \tag{16}$$

$$\chi_n = a^{-n} r^n \sum_{m=0}^{n} P_n^m \left( C_n^m \cos m\phi + \hat{C}_n^m \sin m\phi \right). \tag{17}$$

In equations (8) and (13), $\mathbf{r}$ represents the position vector corresponding to spherical coordinate system moving with the droplet centre. We note from the above representation of the velocity and pressure fields that one has to find the coefficients: $A_n^m$, $B_n^m$, $C_n^m$, $\hat{A}_n^m$, $\hat{B}_n^m$, and $\hat{C}_n^m$ (and similar terms obtained by replacing $n$ by $-n-1$) to solve velocity and pressure fields inside and outside the droplet. In the above expressions, $P_n^m \left( \cos \theta \right)$ is the associated Legendre polynomial.

Before proceeding further, it is convenient to represent the tangential component of the stress vector $\dfrac{\delta}{\mu_o} \left( \mathbf{I} - \mathbf{nn} \right) \cdot \boldsymbol{\sigma}_i \cdot \mathbf{n}$ as $\boldsymbol{\tau}_i$, i.e. the tangential component of $\boldsymbol{\sigma}_i \cdot \mathbf{n}$. Thus, it is now convenient to transform the previously stated boundary conditions (equations (4)-(6)) to obtain the unknown spherical solid harmonics. The transformed sets of boundary conditions are as follows:

$$\mathbf{u}_i^* \cdot \mathbf{e}_r = 0, \tag{18}$$

$$\mathbf{v}_o^* \cdot \mathbf{e}_r = -\mathbf{u}_\infty^* \cdot \mathbf{e}_r, \tag{19}$$

$$r \nabla \cdot \mathbf{u}_i^* - r \nabla \cdot \mathbf{v}_o^* = r \nabla \cdot \mathbf{u}_\infty^* - r \nabla \cdot \boldsymbol{\tau}_i^*, \tag{20}$$

$$\mathbf{r} \cdot \nabla \times \mathbf{u}_i^* - \mathbf{r} \cdot \nabla \times \mathbf{v}_o^* = \mathbf{r} \cdot \nabla \times \mathbf{u}_\infty^* - \mathbf{r} \cdot \nabla \times \boldsymbol{\tau}_i^*, \tag{21}$$

$$\mathbf{r} \cdot \nabla \times \left[ \mathbf{r} \times \left( \boldsymbol{\sigma}_i^* \cdot \mathbf{e}_r \right) \right] - \mathbf{r} \cdot \nabla \times \left[ \mathbf{r} \times \left( \boldsymbol{\sigma}_o^* \cdot \mathbf{e}_r \right) \right] = \mathbf{r} \cdot \nabla \times \left[ \mathbf{r} \times \left( \boldsymbol{\sigma}_\infty^* \cdot \mathbf{e}_r \right) \right], \tag{22}$$

$$\mathbf{r} \cdot \nabla \times \left( \boldsymbol{\sigma}_i^* \cdot \mathbf{e}_r \right) - \mathbf{r} \cdot \nabla \times \left( \boldsymbol{\sigma}_o^* \cdot \mathbf{e}_r \right) = \mathbf{r} \cdot \nabla \times \left( \boldsymbol{\sigma}_\infty^* \cdot \mathbf{e}_r \right), \tag{23}$$

$$\left( \boldsymbol{\sigma}_o^* \cdot \mathbf{e}_r \right) \cdot \mathbf{e}_r + \left( \boldsymbol{\sigma}_\infty^* \cdot \mathbf{e}_r \right) \cdot \mathbf{e}_r - \left( \boldsymbol{\sigma}_i^* \cdot \mathbf{e}_r \right) \cdot \mathbf{e}_r = \Gamma \left( \frac{1}{R_1} + \frac{1}{R_2} \right), \tag{24}$$

where an asterisk over any quantity represents the evaluation of the respective quantity at the undeformed interface ($r = a$), and $\mathbf{e}_r$ represents the radial unit vector. One thing to note here is that at the leading order approximation we calculate velocity and pressure fields by imposing boundary conditions at the undeformed droplet interface (which means $\mathbf{n} = \mathbf{e}_r$) and after that we will calculate the interfacial deformation. Out of the above seven boundary conditions, only the first six (i.e. equations (18)-(23)) are sufficient to determine the velocity and pressure fields $\left( \mathbf{u}_i, p_i, \mathbf{v}_o, p_o \right)$. The seventh boundary condition (i.e. equation (24)) is used later to obtain the deformation of the droplet interface.[47,48]

Similar to the representation of the fields $\mathbf{u}_i$ and $\mathbf{v}_o$, the unperturbed flow field as $r \to \infty$, $\mathbf{u}_\infty$, must also be described in terms of solid spherical harmonics since it also satisfies the Stokes equation. Therefore, we may proceed to write $\mathbf{u}_\infty$ as

$$\mathbf{u}_\infty = \sum_{n=-\infty}^{\infty} \left[ \nabla \times \left( \mathbf{r} \chi_n^\infty \right) + \nabla \Phi_n^\infty + \frac{n+3}{2(n+1)(2n+3)\mu_o} r^2 \nabla p_n^\infty - \frac{n}{(n+1)(2n+3)\mu_o} \mathbf{r} p_n^\infty \right],$$

(25)

where the solid spherical harmonics $\left( p_n^\infty, \Phi_n^\infty \text{ and } \chi_n^\infty \right)$ are defined as:

$$p_n^\infty = \frac{2(2n+3)}{n} \mu_o a^{-n-1} r^n \sum_{m=0}^{n} P_n^m \left( \alpha_n^m \cos m\phi + \hat{\alpha}_n^m \sin m\phi \right),$$

(26)

$$\Phi_n^\infty = \frac{1}{n} a^{-n+1} r^n \sum_{m=0}^{n} P_n^m \left( \beta_n^m \cos m\phi + \hat{\beta}_n^m \sin m\phi \right),$$

(27)

$$\chi_n^\infty = \frac{1}{n(n+1)} a^{-n} r^n \sum_{m=0}^{n} P_n^m \left( \gamma_n^m \cos m\phi + \hat{\gamma}_n^m \sin m\phi \right).$$

(28)

The coefficients $\alpha, \beta$ and $\gamma$ are determined from the knowledge of the imposed flow field as we shall see later. In a similar way, one can represent $\mathbf{V}_\infty$ in terms of solid spherical harmonics $p_n^\infty, \Phi_n^\infty$ and $\chi_n^\infty$ by replacing $\alpha_n^m, \beta_n^m$ and $\gamma_n^m$ by $\zeta_n^m, \eta_n^m$ and $\psi_n^m$ respectively.

Expressing the boundary conditions (equations (18)-(23)) by making use of the above form of the velocity fields, we obtain, through orthogonality of the associated Legendre polynomials,:

$$\sum_{n=1}^{\infty} \left[ \frac{n}{4n+6} A_n^m + n B_n^m \right] = 0,$$

(29)

$$\sum_{n=1}^{\infty} \left[ \left( \frac{n+1}{4n-2} \right) A_{-n-1}^m - (n+1) B_{-n-1}^m \right] = \sum_{n=1}^{\infty} \left[ -\alpha_n^m - \beta_n^m - \alpha_{-n-1}^m - \beta_{-n-1}^m \right],$$

(30)

$$\sum_{n=1}^{\infty} \left[ \frac{n(n+1)}{4n+6} A_n^m + n(n-1) B_n^m + \frac{n(n+1)}{4n-2} A_{-n-1}^m - (n+1)(n+2) B_{-n-1}^m \right] = \sum_{n=1}^{\infty} \left[ (n+1)\alpha_n^m \right.$$
$$\left. - n\alpha_{-n-1}^m + (n-1)\beta_n^m - (n+2)\beta_{-n-1}^m - \delta\lambda \left( 2(n-1)n(n+1) B_n^m + \frac{n^2(n+2)}{2n+3} A_n^m \right) \right],$$

(31)

$$\sum_{n=1}^{\infty} n(n+1) \left( C_n^m - C_{-n-1}^m \right) = \sum_{n=1}^{\infty} \left[ \gamma_n^m + \gamma_{-n-1}^m - \delta\lambda (n-1)(n+1) n C_n^m \right],$$

(32)

$$\sum_{n=1}^{\infty}\left[2n(n+1)(n+2)B_{-n-1}^{m}-\frac{(n+1)^{2}(n-1)}{2n-1}A_{-n-1}^{m}+2\lambda(n-1)(n+1)nB_{n}^{m}\right.$$

$$\left.+\frac{n^{2}(n+2)\lambda}{2n+3}A_{n}^{m}\right]=\sum_{n=1}^{\infty}\left[\begin{array}{c}2\left(n^{2}-1\right)\beta_{n}^{m}+2n(n+2)\beta_{-n-1}^{m}\\+2n(n+2)\alpha_{n}^{m}+2\left(n^{2}-1\right)\alpha_{-n-1}^{m}\end{array}\right], \qquad (33)$$

$$\sum_{n=1}^{\infty}\left[n(n+1)\left(\lambda(n-1)C_{n}^{m}+(n+2)C_{-n-1}^{m}\right)\right]=\sum_{n=1}^{\infty}\left[(n-1)\gamma_{n}^{m}-(n+2)\gamma_{-n-1}^{m}\right], \qquad (34)$$

where $\lambda=\mu_{i}/\mu_{o}$ is the viscosity ratio. Similar equations are obtained for $\hat{A}_{n}^{m},\hat{B}_{n}^{m},\hat{C}_{n}^{m},\hat{A}_{-n-1}^{m},\hat{B}_{-n-1}^{m}$ and $\hat{C}_{-n-1}^{m}$ by replacing $\alpha_{n}^{m},\beta_{n}^{m},\gamma_{n}^{m},\alpha_{-n-1}^{m},\beta_{-n-1}^{m}$ and $\gamma_{-n-1}^{m}$ by $\hat{\alpha}_{n}^{m},\hat{\beta}_{n}^{m},\hat{\gamma}_{n}^{m},\hat{\alpha}_{-n-1}^{m},\hat{\beta}_{-n-1}^{m}$ and $\hat{\gamma}_{-n-1}^{m}$ respectively. After solving equations (29) - (34), we obtain the coefficients $A_{n}^{m},B_{n}^{m},C_{n}^{m}$, etc. of the following form

$$A_{n}^{m}=\frac{(2n+3)}{n\left[(\lambda+1)+\delta\lambda(2n+1)\right]}\left[(2n+3)\alpha_{n}^{m}+(2n-1)\beta_{n}^{m}\right], \qquad (35)$$

$$B_{n}^{m}=-\frac{(2n+3)\alpha_{n}^{m}+(2n-1)\beta_{n}^{m}}{2n\left[(\lambda+1)+\delta\lambda(2n+1)\right]}, \qquad (36)$$

$$C_{n}^{m}=\frac{(1+2n)\left[\left(n^{2}-1+\left(n^{3}+2n^{2}-n-2\right)\delta\right)\lambda+\left(+n^{2}+3n++2\right)\right]}{n\left[\left\{\left(n^{3}+2n^{2}-n-2\right)\delta+\left(n^{2}-1\right)\right\}\lambda+\left(n^{2}+3n+2\right)\right]}\gamma_{n}^{m}, \qquad (37)$$

$$A_{-n-1}^{m}=-\frac{1}{(n+1)\left[(\lambda+1)+\delta\lambda(2n+1)\right]}\left[\left(4n^{2}+4n-3\right)\lambda\alpha_{n}^{m}+\right.$$

$$\left\{\left(4n^{2}+8\delta n^{2}-2\delta-1\right)\lambda+(4n-2)\right\}\beta_{n}^{m}$$

$$\left.+\left\{\left(8n^{2}\delta+4n-2\delta-2\right)\lambda+(4n-2)\right\}\alpha_{-n-1}^{m}\right] \qquad (38)$$

$$B_{-n-1}^{m}=\frac{1}{2\left[\left\{\left(2n^{2}+3n+1\right)\delta+(n+1)\right\}\lambda+(n+1)\right]}\left[\left\{\left(4\delta n+2\delta-2n-1\right)\lambda+2\right\}\alpha_{n}^{m}\right.$$

$$\left.+(1-2n)\lambda\beta_{n}^{m}+\left\{\left(4\delta n+2\delta+2\right)\lambda+2\right\}\beta_{-n-1}^{m}\right], \qquad (39)$$

$$C_{-n-1}^{m}=\frac{\left\{\left(\delta n^{2}-2\delta n+\delta-n+1\right)\lambda+(n-1)\right\}\gamma_{n}^{m}+\left\{\left(-\delta n^{2}-\delta n+2\delta-n+1\right)\lambda-(n+2)\right\}\gamma_{-n-1}^{m}}{n(n+1)\left\{\left(\delta n^{2}+\delta n-2\delta+n-1\right)\lambda+(n+2)\right\}}. \qquad (40)$$

The coefficients $\hat{A}_{n}^{m},\hat{B}_{n}^{m},\hat{C}_{n}^{m}$, etc. can be obtained by replacing $\alpha_{n}^{m},\beta_{n}^{m},\gamma_{n}^{m}$, etc. by $\hat{\alpha}_{n}^{m},\hat{\beta}_{n}^{m},\hat{\gamma}_{n}^{m}$, etc.

We now proceed to represent the parameters $\alpha_n^m, \beta_n^m, \gamma_n^m, \hat{\alpha}_n^m, \hat{\beta}_n^m$ and $\hat{\gamma}_n^m$ in terms of the known imposed flow field and yet-to-be-determined droplet velocity. The droplet velocity is eventually found out by asserting that the droplet is in a force-free condition. We reiterate that $\mathbf{u}_\infty = \mathbf{V}_\infty - \mathbf{U}_d$. Upon multiplying this by the unit radial vector, we obtain

$$\mathbf{u}_\infty \cdot \mathbf{e}_r = \mathbf{V}_\infty \cdot \mathbf{e}_r - \left[ U_{d,x} \cos\phi P_1^1(\cos\theta) + U_{d,y} \sin\phi P_1^1(\cos\theta) + U_{d,z} P_1^0(\cos\theta) \right], \quad (41)$$

where $U_{d,x}$, $U_{d,y}$ and $U_{d,z}$ are x-, y- and z- component of droplet velocity respectively.

The left hand side of equation (41) can be written as

$$\mathbf{u}_\infty \cdot \mathbf{e}_r = \sum_{n=1}^\infty \sum_{m=0}^n \left[ \alpha_n^m \left(\frac{r}{a}\right)^{n+1} + \beta_n^m \left(\frac{r}{a}\right)^{n-1} + \alpha_{-n-1}^m \left(\frac{r}{a}\right)^{-n} + \beta_{-n-1}^m \left(\frac{r}{a}\right)^{n-2} \right] P_n^m \cos m\phi$$
$$+ \sum_{n=1}^\infty \sum_{m=0}^n \left[ \hat{\alpha}_n^m \left(\frac{r}{a}\right)^{n+1} + \hat{\beta}_n^m \left(\frac{r}{a}\right)^{n-1} + \hat{\alpha}_{-n-1}^m \left(\frac{r}{a}\right)^{-n} + \hat{\beta}_{-n-1}^m \left(\frac{r}{a}\right)^{n-2} \right] P_n^m \sin m\phi$$

$$(42)$$

Similarly, the $\mathbf{r} \cdot \nabla \times$ operator on $\mathbf{u}_\infty$ yields

$$\mathbf{r} \cdot \nabla \times \mathbf{u}_\infty = \sum_{n=1}^\infty \sum_{m=0}^n \left[ \gamma_n^m \left(\frac{r}{a}\right)^n + \gamma_{-n-1}^m \left(\frac{r}{a}\right)^{-n-1} \right] P_n^m \cos m\phi$$
$$+ \sum_{n=1}^\infty \sum_{m=0}^n \left[ \hat{\gamma}_n^m \left(\frac{r}{a}\right)^n + \hat{\gamma}_{-n-1}^m \left(\frac{r}{a}\right)^{-n-1} \right] P_n^m \sin m\phi$$

$$(43)$$

Similarly, the $\mathbf{V}_\infty$ may be written in analogy to equation (42).

$$\mathbf{V}_\infty \cdot \mathbf{e}_r = \sum_{n=1}^\infty \sum_{m=0}^n \left[ \zeta_n^m \left(\frac{r}{a}\right)^{n+1} + \eta_n^m \left(\frac{r}{a}\right)^{n-1} + \zeta_{-n-1}^m \left(\frac{r}{a}\right)^{-n} + \eta_{-n-1}^m \left(\frac{r}{a}\right)^{n-2} \right] P_n^m \cos m\phi$$
$$+ \sum_{n=1}^\infty \sum_{m=0}^n \left[ \hat{\zeta}_n^m \left(\frac{r}{a}\right)^{n+1} + \hat{\eta}_n^m \left(\frac{r}{a}\right)^{n-1} + \hat{\zeta}_{-n-1}^m \left(\frac{r}{a}\right)^{-n} + \hat{\eta}_{-n-1}^m \left(\frac{r}{a}\right)^{n-2} \right] P_n^m \sin m\phi ,$$

$$(44)$$

and in line with equation (43), we may write

$$\mathbf{r} \cdot \nabla \times \mathbf{V}_\infty = \sum_{n=1}^\infty \sum_{m=0}^n \left[ \psi_n^m \left(\frac{r}{a}\right)^n + \psi_{-n-1}^m \left(\frac{r}{a}\right)^{-n-1} \right] P_n^m \cos m\phi$$
$$+ \sum_{n=1}^\infty \sum_{m=0}^n \left[ \hat{\psi}_n^m \left(\frac{r}{a}\right)^n + \hat{\psi}_{-n-1}^m \left(\frac{r}{a}\right)^{-n-1} \right] P_n^m \sin m\phi ,$$

$$(45)$$

where $\zeta, \eta$ and $\psi$ are the coefficients which define the solid spherical harmonics related to $\mathbf{V}_\infty$ in the following form

$$p_n^\infty = \frac{2(2n+3)}{n} \mu_o a^{-n-1} r^n \sum_{m=0}^n P_n^m \left( \zeta_n^m \cos m\phi + \hat{\zeta}_n^m \sin m\phi \right)$$

$$\Phi_n^\infty = \frac{1}{n} a^{-n+1} r^n \sum_{m=0}^n P_n^m \left( \eta_n^m \cos m\phi + \hat{\eta}_n^m \sin m\phi \right) \qquad . \qquad (46)$$

$$\chi_n^\infty = \frac{1}{n(n+1)} a^{-n} r^n \sum_{m=0}^n P_n^m \left( \psi_n^m \cos m\phi + \hat{\psi}_n^m \sin m\phi \right)$$

From equation (41), (42) and (44), we can write

$$\beta_1^0 = \eta_1^0 - U_{d,z}, \quad \beta_1^1 = \eta_1^1 - U_{d,x}, \quad \hat{\beta}_1^1 = \hat{\eta}_1^1 - U_{d,y} \qquad (47)$$

while for all other values of $n$ and $m$

$$\left\{ \alpha_n^m, \alpha_{-n-1}^m, \hat{\alpha}_n^m, \hat{\alpha}_{-n-1}^m \right\} = \left\{ \zeta_n^m, \zeta_{-n-1}^m, \hat{\zeta}_n^m, \hat{\zeta}_{-n-1}^m \right\},$$

$$\left\{ \beta_n^m, \beta_{-n-1}^m, \hat{\beta}_n^m, \hat{\beta}_{-n-1}^m \right\} = \left\{ \eta_n^m, \eta_{-n-1}^m, \hat{\eta}_n^m, \hat{\eta}_{-n-1}^m \right\}. \qquad (48)$$

From equations (43) and (45), and making use of the fact that the droplet velocity is constant, we can easily show that $\left\{ \gamma_n^m, \gamma_{-n-1}^m, \hat{\gamma}_n^m, \hat{\gamma}_{-n-1}^m \right\} = \left\{ \psi_n^m, \psi_{-n-1}^m, \hat{\psi}_n^m, \hat{\psi}_{-n-1}^m \right\}.$

## B. Velocity of droplet

The velocity of a droplet can be obtained from the force free condition of the form

$$\frac{4\pi}{3} a^3 (\rho_i - \rho_o) g \mathbf{e}_z - 4\pi \nabla \left( r^3 p_{-2} \right) = 0, \qquad (49)$$

where $\mathbf{e}_z$ is the unit vector along z-coordinate which is taken as the direction of gravity. The solid spherical harmonic $p_{-2}$ is of the form

$$p_{-2} = \mu_o a r^{-2} \left\{ A_{-2}^0 P_1 (\cos\theta) + A_{-2}^1 \cos\phi P_1^1 (\cos\theta) + \hat{A}_{-2}^1 \sin\phi P_1^1 (\cos\theta) \right\}. \qquad (50)$$

Now substituting equations (50) and the expressions of $A_{-2}^0, A_{-2}^1$ and $\hat{A}_{-2}^1$ in equation (49), we obtain

$$\left[ \frac{\left\{ 5\alpha_1^0 + (3+6\delta)\beta_1^0 \right\} \lambda + 2\beta_1^0}{(3\delta\lambda + \lambda + 1)} \right] = -\frac{2a^2 g}{3\mu_o} (\rho_i - \rho_o), \qquad (51)$$

$$\left[ \frac{\left\{ 5\alpha_{1,1} + \left(3+6\delta\right)\beta_1^1 \right\}\lambda + 2\beta_1^1}{\left(3\delta\lambda + \lambda + 1\right)} \right] = 0, \tag{52}$$

$$\left[ \frac{\left\{ 5\hat{\alpha}_1^1 + \left(3+6\delta\right)\hat{\beta}_1^1 \right\}\lambda + 2\hat{\beta}_1^1}{\left(3\delta\lambda + \lambda + 1\right)} \right] = 0. \tag{53}$$

To arrive these relations we have used the fact that $\alpha_{-2}^0 = \alpha_{-2}^1 = \hat{\alpha}_{-2}^1 = 0$ because force due to the unperturbed pressure field $\nabla\left(r^3 p_{-2}^\infty\right) = 0$ [48]. Now we substitute $\alpha$ and $\beta$ using expressions of equations (47) and (48) in terms of $\zeta$ and $\eta$. Finally, we obtain the expressions of droplet velocity of the form

$$U_{d,x} = \eta_1^1 + \left( \frac{5\lambda}{2+3\lambda+6\delta\lambda} \right)\zeta_1^1, \tag{54}$$

$$U_{d,y} = \hat{\eta}_1^0 + \left( \frac{5\lambda}{2+3\lambda+6\delta\lambda} \right)\hat{\zeta}_1^1, \tag{55}$$

$$U_{d,z} = \frac{2\left(\rho_i - \rho_o\right)a^2 g}{3\mu_o}\left( \frac{1+\lambda+3\delta\lambda}{2+3\lambda+6\delta\lambda} \right) + \eta_1^0 + \left( \frac{5\lambda}{2+3\lambda+6\delta\lambda} \right)\zeta_1^0. \tag{56}$$

This form of the droplet velocity can be represented in terms of the imposed velocity field evaluated at the droplet centre in the following way. The imposed velocity and pressure fields at the droplet centre can be represented by

$$\left[\mathbf{V}_\infty\right]_{r=0} = \left[\nabla\Phi_1^\infty\right]_{r=0} = \eta_1^1\mathbf{e}_x + \hat{\eta}_1^1\mathbf{e}_y + \eta_1^0\mathbf{e}_z \tag{57}$$

and

$$\left[\nabla p_\infty\right]_{r=0} = \left[\mu_o \nabla^2 \mathbf{V}_\infty\right]_{r=0} = \frac{10\mu_o}{a^2}\left( \zeta_1^1\mathbf{e}_x + \hat{\zeta}_1^1\mathbf{e}_y + \zeta_1^0\mathbf{e}_z \right). \tag{58}$$

Substituting (57) and (58) in equations (54)-(56), we obtain the modified Faxén law for a droplet in the presence of interfacial slip of the form

$$\mathbf{U}_d = \frac{2\left(\rho_i - \rho_o\right)a^2 g}{3\mu_o}\left( \frac{1+\lambda+3\delta\lambda}{2+3\lambda+6\delta\lambda} \right)\mathbf{e}_z + \left[\mathbf{V}_\infty\right]_{r=0} + \frac{a^2\lambda}{2\left(2+3\lambda+6\delta\lambda\right)}\left[\nabla^2\mathbf{V}_\infty\right]_{r=0}. \tag{59}$$

In the limit of viscosity ratio $\lambda \to \infty$ (the case of solid sphere), from equation (59) one can easily obtain the velocity of a sedimenting solid sphere in the presence of interfacial slip; i.e. the classical result obtained by Basset of the form $\mathbf{U}_d = \left( \frac{1+3\delta/a}{1+2\delta/a} \right)\frac{\mathbf{F}}{6\pi\mu_o a}$, [31] where $\mathbf{F}$ is the force acting on the particle, as $\frac{4\pi}{3}a^3\left(\rho_i - \rho_o\right)g\mathbf{e}_z$.

## C. Deformation of droplet

The interface, in the case of small Capillary number may be denoted as $r = a\left(1 + f\left(\theta, \phi\right)\right)$. We appeal to the normal stress balance at the undeformed interface to evaluated the equation of the deformed interface. Mathematically, we have

$$\left(\sigma_o^* \cdot \mathbf{e}_r\right) \cdot \mathbf{e}_r + \left(\sigma_\infty^* \cdot \mathbf{e}_r\right) \cdot \mathbf{e}_r - \left(\sigma_i^* \cdot \mathbf{e}_r\right) \cdot \mathbf{e}_r = \Gamma\left(\frac{1}{R_1} + \frac{1}{R_2}\right), \tag{60}$$

where the curvature of the deformed interface can be given by [47,48] $\left(\frac{1}{R_1} + \frac{1}{R_2}\right) = \frac{2}{a} - \frac{2f}{a} - \frac{1}{a}\left[\frac{1}{\sin^2\theta}\frac{\partial^2 f}{\partial\phi^2} + \frac{1}{\sin\theta}\frac{\partial}{\partial\theta}\left(\sin\theta\frac{\partial f}{\partial\theta}\right)\right]$. The function $f\left(\theta, \phi\right)$ can be regarded as a deviation function, and our analysis is valid for small values of $f$. One can split $f$ in terms of the spherical surface harmonics as

$$f\left(\theta, \phi\right) = \sum_{n=1}^{\infty} L_n S_n, \tag{61}$$

where $L_n S_n$ denotes $\sum_{m=0}^{n}\left(L_n^m \cos m\phi + \hat{L}_n^m \sin m\phi\right) P_n^m$; $S_n$ depicting the spherical surface harmonic. Using this form, we can write the curvature as $\left(\frac{1}{R_1} + \frac{1}{R_2}\right) = \frac{1}{a}\left[2 + \sum_{n=1}^{\infty}\left(n\left(n+1\right) - 2\right) L_n S_n\right]$. Using this form of the curvature and the form of the stresses as shown in Appendix A, we obtain

$$
\begin{aligned}
\sum_{n=1}^{\infty}&\left[2n\left(n-1\right)\frac{\lambda}{a}\Phi_n^* + \frac{\left(n^2 - n - 3\right)\lambda a}{\left(2n+3\right)\mu_i}p_n^* - \frac{2\left(n+1\right)\left(n+2\right)}{a}\Phi_{-n-1}^* \right.\\
&+ \frac{\left(n^2 + 3n - 1\right)a}{\left(2n-1\right)\mu_o}p_{-n-1}^* - \frac{2n\left(n-1\right)}{a}\Phi_n^{\infty*} - \frac{\left(n^2 - n - 3\right)a}{\left(2n+3\right)\mu_o}p_n^{\infty*}\\
&\left. - \frac{2\left(n+1\right)\left(n+2\right)}{a}\Phi_{-n-1}^{\infty*} + \frac{\left(n^2 + 3n - 1\right)a}{\left(2n-1\right)\mu_o}p_{-n-1}^{\infty*}\right] - \frac{a}{\mu_o}\left(p_0^\infty + p_{-1}^\infty\right)\\
&- \frac{\left(\rho_i - \rho_o\right)ga}{\mu_o}P_1^0\left(\cos\theta\right) = -\frac{\Gamma}{\mu_o}\left(2 + \sum_{n=1}^{\infty}\left(n\left(n+1\right) - 2\right) L_n S_n\right)
\end{aligned}
\tag{62}
$$

For $n = 0$, equation (62) yields $-\frac{a}{\mu_o}\left(p_0^\infty + p_{-1}^\infty\right) = -\frac{2\Gamma}{\mu_o}$. For $n=1$, one finds zero deformation: classical Hadamard-Rybcznski solution of a sedimenting droplet.[46] And for $n \geq 2$, we obtain

$$L_n^m = \frac{\mu_o}{\Gamma} \frac{1}{\left(n^2 + n - 2\right)(n+1)n\left\{\delta\lambda\left(2n+1\right) + \lambda + 1\right\}}$$
$$\times$$
$$\left[ \alpha_n^m \left\{ \left(\left(16n^3 + 8n^4 - 2n^2 - 16n - 6\right)\delta + 4n^3 + 6n^2 + 2n + 3\right)\lambda + \left(4n^3 + 6n^2 - 4n - 6\right)\right\} \right.$$
$$\left. + \beta_n^m \left\{ \left(\left(8n^4 + 16n^3 - 2n^2 - 4n\right)\delta + 4n^3 + 6n^2 + 2n - 3\right)\lambda + \left(4n^3 + 6n^2 - 4n\right)\right\} \right].$$

$$(63)$$

Identical expression is obtained for $\hat{L}_n^m$ with $\alpha_n^m$ and $\beta_n^m$ in equation (63) replaced by $\hat{\alpha}_n^m$ and $\hat{\beta}_n^m$ respectively.

## IV. ILLUSTRATION FOR COMMON IMPOSED FLOW FIELDS

### A. Shear flow

At first we consider a density matched droplet suspending in a shear flow of the form

$$\mathbf{V}_\infty = G\left(y + l\right)\mathbf{e}_x,$$

$$(64)$$

where $G$ is the shear rate and $l$ is the distance between the droplet center and the axis of zero velocity. To determine the coefficients $\zeta, \eta$ and $\psi$ from (64), we calculate the following expressions

$$\mathbf{V}_\infty \cdot \mathbf{e}_r = Gl\sin\theta\cos\phi + Gr\sin^2\theta\sin\phi\cos\phi = Gl\cos\phi P_1^1 + \frac{Gr}{6}\sin 2\phi P_2^2$$

$$(65)$$

and

$$\mathbf{r} \cdot \nabla \times \mathbf{V}_\infty = -Gr\cos\theta = -GrP_1^0.$$

$$(66)$$

Comparing equations (65) and (66) with equations (44) and (45) respectively, we obtain

$$\eta_1^1 = Gl,$$
$$\hat{\eta}_2^2 = \frac{Ga}{6},$$
$$\psi_1^0 = -Ga,$$

$$(67)$$

and all other coefficients are zero. Substituting these coefficients in equations (54)-(56) results in

$$U_{d,x} = Gl,$$
$$U_{d,y} = U_{d,z} = 0.$$
(68)

The interface deformation is obtained from (63) of the form

$$\hat{L}_2^2 = \frac{\mu_o G a}{\Gamma} \left[ \frac{\lambda(80\delta + 19) + 16}{48(\lambda(5\delta + 1) + 1)} \right].$$
(69)

Consequently, the equation of the deformed interface is obtained as:

$$r = a\left(1 + \hat{L}_2^2 \sin 2\phi P_2^2\right) = a\left[1 + \frac{\mu_o G a}{\Gamma}\left\{\frac{\lambda(80\delta + 19) + 16}{48(\lambda(5\delta + 1) + 1)}\right\}\sin 2\phi P_2^2\right].$$
(70)

By setting $\delta = 0$, we obtain the classical result of Taylor.[6,7] Equation (70) has previously been obtained by Ramchandran and Leal at the leading order approximation of Capillary number.[37]

## B. Poiseuille flow

Now we consider the effect of interfacial slip for the case of a density matched droplet suspending in a unbounded cylindrical Poiseuille flow. The droplet centre is at a distance $b$ from the centre of maximum velocity ($U_0$). The imposed velocity profile with respect to the droplet centre is[44]

$$\mathbf{V}_\infty = U_0\left[1 - \left(\frac{r}{R_0}\right)^2 \sin^2\theta - \left(\frac{b}{R_0}\right)^2 - \frac{2rb}{R_0^2}\sin\theta\cos\phi\right]\mathbf{e}_z,$$
(71)

where $R_0$ is the distance from the maximum velocity to the point of zero velocity. The coefficients $\zeta, \eta$ and $\psi$ related to $\mathbf{V}_\infty$ are obtained by using equations (44) and (45) in the following way:

$$\mathbf{V}_\infty \cdot \mathbf{e}_r = U_0\left[1 - \left(\frac{b}{R_0}\right)^2 - \frac{2}{5}\left(\frac{r}{R_0}\right)^2\right]P_1^0 - \frac{2}{3}\frac{br}{R_0^2}\cos\phi P_2^1 + \frac{2}{5}\left(\frac{r}{R_0}\right)^2 P_3^0$$
(72)

and

$$\mathbf{r} \cdot \nabla \times \mathbf{V}_\infty = U_0\frac{2br}{R_0^2}\sin\phi P_1^1.$$
(73)

From these equations we can identify the following non-zero coefficients

$$\zeta_1^0 = -\frac{2}{5}U_0\left(\frac{a}{R_0}\right)^2, \ \eta_1^0 = U_0\left(1-\left(\frac{b}{R_0}\right)^2\right), \ \eta_2^1 = -\frac{2}{3}\frac{abU_0}{R_0^2}, \ \eta_3^0 = \frac{2}{5}U_0\left(\frac{a}{R_0}\right)^2, \ \hat{\psi}_1^1 = \frac{2ab}{R_0^2}U_0,$$

$$(74)$$

and all other coefficients are zero. From equation (74), we conclude that the velocity fields will contain the following coefficients: $A_1^0, B_1^0, \hat{C}_1^1, A_2^1, B_2^1, A_3^0, B_3^0, A_{-2}^0, B_{-2}^0, A_{-3}^1, B_{-3}^1, A_{-4}^0$ and $B_{-4}^0$.

Now to obtain the velocity of the droplet we use the expressions of (74) in the equations (54)-(56) and obtain

$$U_{d,x} = U_{d,y} = 0,$$

$$U_{d,z} = U_0\left[1-\left(\frac{b}{R_0}\right)^2 - \frac{2\lambda}{2+(3+6\delta)\lambda}\left(\frac{a}{R_0}\right)^2\right].$$

$$(75)$$

In the absence of interfacial slip ($\delta=0$), the above relations reduce to the results of Hetsroni and Haber.[48] The interface deformation is obtained from (63) of the form:

$$\hat{L}_2^1 = -\frac{\mu_o U_0}{\Gamma}\frac{ab}{R_0^2}\left[\frac{(19+80\delta)\lambda+16}{12(1+5\delta)\lambda+12}\right],$$

$$L_3^0 = \frac{\mu_o U_0}{\Gamma}\left(\frac{a}{R_0}\right)^2\left[\frac{(11+70\delta)\lambda+10}{20(1+7\delta)\lambda+20}\right].$$

$$(76)$$

Therefore, the equation of the deformed interface is written as:

$$r = a\left(1+L_2^1\cos\phi P_2^1 + L_3^0 P_3^0\right)$$

$$= a\left[1-\frac{\mu_o U_0}{\Gamma}\frac{ab}{R_0^2}\left[\frac{(19+80\delta)\lambda+16}{12(1+5\delta)\lambda+12}\right]\cos\phi P_2^1 + \frac{\mu_o U_0}{\Gamma}\left(\frac{a}{R_0}\right)^2\left[\frac{(11+70\delta)\lambda+10}{20(1+7\delta)\lambda+20}\right]P_3^0\right].$$

$$(77)$$

It is to be noted that a major assumption in arriving at the above expressions is that the imposed flow is unaffected by the motion of the particle; incorporating the wall effects involves the method of reflections which we do not pursue here. [49]

## V. SUMMARY AND DISCUSSION

It is apparent from equation (75) that velocity of the drop is less in the presence of interfacial slip $(\delta>0)$ as compared to the case of no-slip $(\delta=0)$. By equating the drop velocity in the presence of interfacial slip (equation (75)) with the drop velocity without slip but having different viscosity, we obtain an equivalent viscosity ratio of the drop in the following form

$$\mathbf{U}_d\big|_{\lambda=\lambda_{eq},\delta=0} = \mathbf{U}_d\big|_{\delta\neq0}$$

$$\Rightarrow \lambda_{eq} = \frac{\lambda}{1+3\lambda\delta}. \tag{78}$$

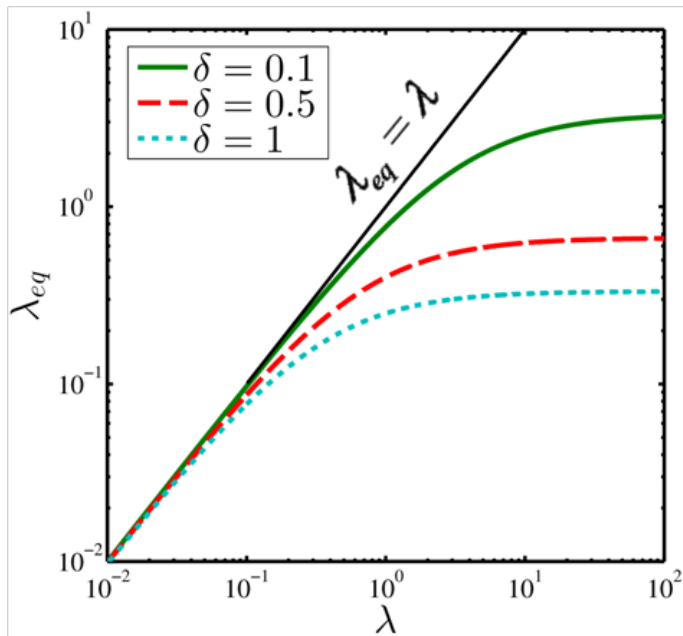

FIG. 2. Variation of equivalent viscosity ratio $\left(\lambda_{eq}\right)$ with the actual viscosity ratio $\left(\lambda\right)$ for three different values of slip parameter $\left(\delta\right)$.

The variation of the equivalent viscosity $\lambda_{eq}$ with the actual viscosity ratio is shown in Figure 2 for three different values of slip parameter $\left(\delta\right)$. The black straight line represents the $\lambda_{eq} = \lambda$ line. Figure 2 depicts that the equivalent viscosity ratio $\lambda_{eq}$ is less as compared to the actual viscosity ratio $\lambda$, and this decrease in effective viscosity ratio is very significant for higher values of $\lambda$. So, the hydrodynamic slip at the fluid-fluid interface results in making the drop less viscous, apparently.

To summarize, a general derivation of the velocity and pressure field is obtained from fundamental considerations of spherical harmonic solutions for a droplet in a suspending medium with any arbitrary (Stokesian) imposed flow field, with due consideration of a interfacial slip condition. The results indicate corrections to the velocity field and the concomitant interface deformation at the leading order of Capillary number. A general expression for the droplet velocity is also obtained as a function of the ratio of the two fluid viscosities and the slip parameter. The results of Taylor[6] and Basset[31] are obtained from the general formulae. Also, we obtain the deformation, which was obtained using vector harmonics by Ramachandran and Leal,[37] for a simple shear flow of a droplet with slip. We then proceed to obtain the deformation of the droplet with interfacial slip in a Poiseuille flow.

## Appendix A. Boundary conditions in terms of spherical harmonics

The radial velocity boundary condition in terms of spherical harmonics is given as

$$\mathbf{v} \cdot \mathbf{e}_r = \sum_{n=-\infty}^{\infty} \left[ \frac{n}{2\mu(2n+3)} r p_n + \frac{n}{r} \Phi_n \right] \tag{A.1}$$

Similarly, the two more boundary conditions which yield simplifying relationships between the harmonics are obtained as [44]

$$-r\nabla \cdot \mathbf{v} = \sum_{n=-\infty}^{\infty} \left[ \frac{n(n+1)}{2\mu(2n+3)} r p_n + \frac{n(n-1)}{r} \Phi_n \right] \tag{A.2}$$

$$\mathbf{r} \cdot \nabla \times \mathbf{v} = \sum_{n=-\infty}^{\infty} n(n+1) \chi_n \tag{A.3}$$

In the case of the slip boundary condition at the interface, it is imperative to also obtain relations for stress in terms of the spherical harmonics [48]. These relations which are used in section 2 are listed below

$$-r\nabla \cdot \boldsymbol{\tau}_i = \frac{\delta\lambda}{r} \sum_{n=1}^{\infty} \left[ \frac{2(n-1)n(n+1)}{r} \Phi_n + \frac{n^2(n+2)}{\mu_i(2n+3)} r p_n \right], \tag{A.4}$$

$$\mathbf{r} \cdot \nabla \times \boldsymbol{\tau}_i = \frac{\delta\lambda}{r} \sum_{n=1}^{\infty} n(n^2-1) \chi_n, \tag{A.5}$$

$$\mathbf{r} \cdot \nabla \times (\boldsymbol{\sigma} \cdot \mathbf{e}_r) = \frac{\mu}{r} \sum_{n=-\infty}^{\infty} (n-1)n(n+1) \chi_n, \tag{A.6}$$

$$\mathbf{r} \cdot \nabla \times (\mathbf{r} \times (\boldsymbol{\sigma} \cdot \mathbf{e}_r)) = -\mu \sum_{n=-\infty}^{\infty} \left[ \frac{2(n-1)n(n+1)}{r} \Phi_n + \frac{n^2(n+2)}{\mu(2n+3)} r p_n \right], \tag{A.7}$$

$$(\boldsymbol{\sigma} \cdot \mathbf{e}_r) \cdot \mathbf{e}_r = \frac{\mu}{r} \sum_{n=-\infty}^{\infty} \left[ \frac{2n(n-1)}{r} \Phi_n + \frac{n^2-n-3}{\mu(2n+3)} r p_n \right]. \tag{A.8}$$